\newif\ifdraft
\draftfalse
\drafttrue

\ifdraft
\documentclass[preprint,onecolumn,showpacs,amsmath,amssymb,aps]{revtex4-1}
\else
\documentclass[twocolumn,showpacs,amsmath,amssymb,aps]{revtex4-1}
\fi

\usepackage{graphicx}
\usepackage{dcolumn}
\usepackage{bm}
\usepackage{amsmath}
\usepackage{amssymb}
\usepackage{url}
\usepackage{subfigure}
\usepackage{algorithm}
\usepackage{algorithmic}
\usepackage{enumerate}
\usepackage{slashbox}
\usepackage{setspace}

\usepackage[usenames,dvipsnames,svgnames,table]{xcolor}
\definecolor{light-gray}{gray}{0.8}

\begin{document}

\preprint{APS/123-QED}

\title{Link community detection through global optimization and the inverse resolution limit of partition density}

\author{Juyong Lee}
\email{juyong.lee@nih.gov}
\affiliation{Laboratory of Computational Biology, National Heart, Lung, and Blood Institute (NHLBI), National Institutes of Health (NIH), Bethesda, Maryland 20892, U.S.A.}

\author{Zhong-Yuan Zhang}
\email{zhyuanzh@gmail.com}
\affiliation{School of Statistics, Central University of Finance and Economics, Beijing, China}
\thanks{\\ The first two authors contributed equally to this paper.\\}


\author{Jooyoung Lee}
\affiliation{School of Computational Sciences, Korea Institute of Advanced Study, Seoul, Korea}

\author{Bernard R. Brooks}
\affiliation{Laboratory of Computational Biology, National Heart, Lung, and Blood Institute (NHLBI), National Institutes of Health (NIH), Bethesda, Maryland 20892, USA}

\author{Yong-Yeol Ahn} %
\affiliation{School of Informatics and Computing, Indiana University Bloomington}%

\date{\today}

\begin{abstract}

We investigate the possibility of global optimization-based overlapping community
detection, using link community framework. We first show that partition
density, the original quality function used in link community detection method,
is not suitable as a quality function for global optimization because it
prefers breaking communities into triangles except in highly limited
conditions. We analytically derive those conditions and confirm it with
computational results on direct optimization of various synthetic and
real-world networks. To overcome this limitation, we propose alternative
approaches combining the weighted line graph transformation and existing quality
functions for node-based communities. We suggest a new line graph weighting 
scheme, a normalized Jaccard index. Computational results show that 
community detection using the weighted line graphs generated with 
the normalized Jaccard index leads to a more accurate community structure.


\end{abstract}

\maketitle

\section{\label{Introduction}Introduction}

Finding community structure is essential in understanding meso-scale
organization of complex networks. Most commonly, community detection is
performed by assigning nodes into groups that optimizes a quality function,
which measures how meaningful the grouping is~\cite{Fortunato2010}.  Community
detection methods can be classified into two main categories based on whether
they allow a node to be included in multiple communities (overlapping
communities) or not (disjoint communities). For the latter, the most widely
used quality measure is modularity~\cite{Newman2004}. It measures, for each
community, the difference between the number of links between the nodes in the
same community and the expected number of links when the network is randomly re-wired.
Various optimization methods have been applied to optimize
modularity~\cite{Fortunato2010, lee2012modularity, zhang2009modularity,
Bagrow2012}. Although modularity has been widely applied to analyze various
social and biological networks~\cite{Lee2013a, Lee2013b}, several drawbacks
have been found~\cite{Fortunato2010,good2010performance}. 
One of the most important problem is so-called ``resolution
limit''~\cite{fortunato2007resolution}.  As a network becomes larger, the
expected number of links within a group decrease, eventually leading to the
situation where even merging two distinct complete cliques is better than
keeping them separated. Thus, small but meaningful communities in a large
network may not be detectable with modularity.

Meanwhile, it has been argued that communities overlap pervasively in many
real-world networks~\cite{Palla2005, ahn2010link}.  For example, in social
networks, each person participates in multiple social groups.  In biological
networks, a protein may play diverse roles in multiple biological
processes~\cite{Lee2011,Lee2013a,Lee2013b,Malik2014a}.  
Among many overlapping community
detection methods that have been suggested~\cite{Palla2005, Lancichinetti2009,
ahn2010link, Evans2010, Esquivel2011, Lancichinetti2011a, Yang2012, Xie2013a,
Gopalan2013, zhang2013overlapping}, here we focus on the ``link community''
paradigm, where the communities are redefined as sets of links (edges) rather
than nodes~\cite{ahn2010link,Evans2010}. This framework provides a clean way to
handle pervasive overlaps between communities because identifying communities
of \emph{links} in the original graph is equivalent to identifying disjoint
communities of \emph{nodes} in the ``line graph'' of the original
graph~\cite{west2001introduction, Evans2009, Evans2010, ahn2010link}. 

\begin{figure}[ht!]
\includegraphics[width=0.85\textwidth]{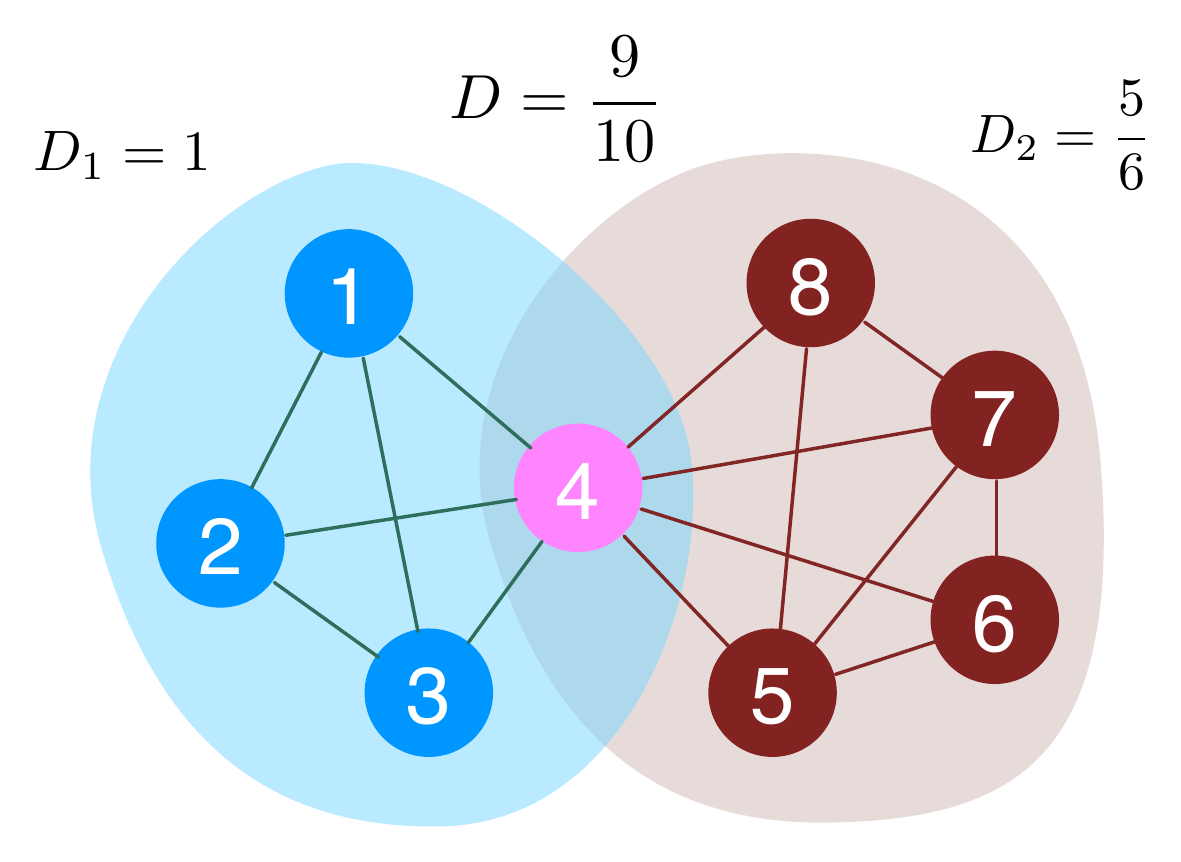}
\caption{A toy example that shows how partition density is calculated.
The local partition density of the blue nodes $D_1$ is one because it is a clique, while that of the red nodes $D_2$ is less than one.
The total partition density $D$ of the community structure is the weighted sum of two local partition densities, 0.9.}
\label{Fig:01}
\end{figure}

As modularity is not well-defined for the link-groupings, ``partition density''
was proposed as a quality function for link communities~\cite{ahn2010link}.
For an undirected and unweighted network, imagine a disjoint partition of links
$C = \{ C_1 , \dots ,C_{n_c} \} $ where $n_c$ is the number of link
communities.  The local partition density of a link community $C_\alpha$ is:
\begin{equation} \label{eq:local_partition_density} D_\alpha =
\frac{m_\alpha-\underline{m}_\alpha}{\overline{m}_\alpha-\underline{m}_\alpha},
\end{equation}
where $m_\alpha$ is the number of links in the community $C_\alpha$,
$\underline{m}_\alpha=(n_\alpha-1)$ and
$\displaystyle\overline{m}_\alpha=\frac{n_\alpha(n_\alpha-1)}{2}$ are the
minimum and maximum possible numbers of links between the induced nodes that
the links in $C_\alpha$ touch, assuming that the nodes in $C_\alpha$ are
connected, and $n_\alpha$ is the number of the induced nodes.  If the induced
nodes are not connected, $D_\alpha$ is set to $0$.  The
partition density of the network is:
\begin{equation} \label{eq:total_partition_density} D=\sum_{\alpha=1}^{n_c}
\frac{m_\alpha}{M} D_\alpha, \end{equation}
where $M$ is the number of links in the network.  Figure~\ref{Fig:01} shows a
toy example that illustrates how partition density is calculated.  By employing
hierarchical clustering and Jaccard index-based link similarity measure, a previous
study argued that partition density can be used to identify meaningful
communities evaluated by the similarity of the metadata of the
nodes~\cite{ahn2010link}.  Additionally, as partition density only uses local
information, it was suggested that partition density is free from the problem
of resolution limit observed in modularity~\cite{fortunato2007resolution,
ahn2010link}. 

Here we show, while partition density may not have the problem of the
resolution limit, that partition density suffers from an inverse
resolution limit: a strong preference towards small cliques, 
due to the fact that partition density is a simple density measure
that does not take any null model into account.  We show that direct
optimization of partition density simply identifies most 3-cliques
(triangle) in the network. We analyze when exactly triangles are 
favored or not by using toy models and a systematic
classification of triangles based on their connectivity.
Our analysis demonstrates that larger communities are only 
favored in highly limited conditions.
We then explore alternative direct-optimization approaches---a
combination of weighted line graph transformation and other quality
functions---to identify overlapping communities through the link
community framework.

\section{Preference to triangles}

\begin{figure}[ht!]
  \includegraphics[width=0.85\columnwidth]{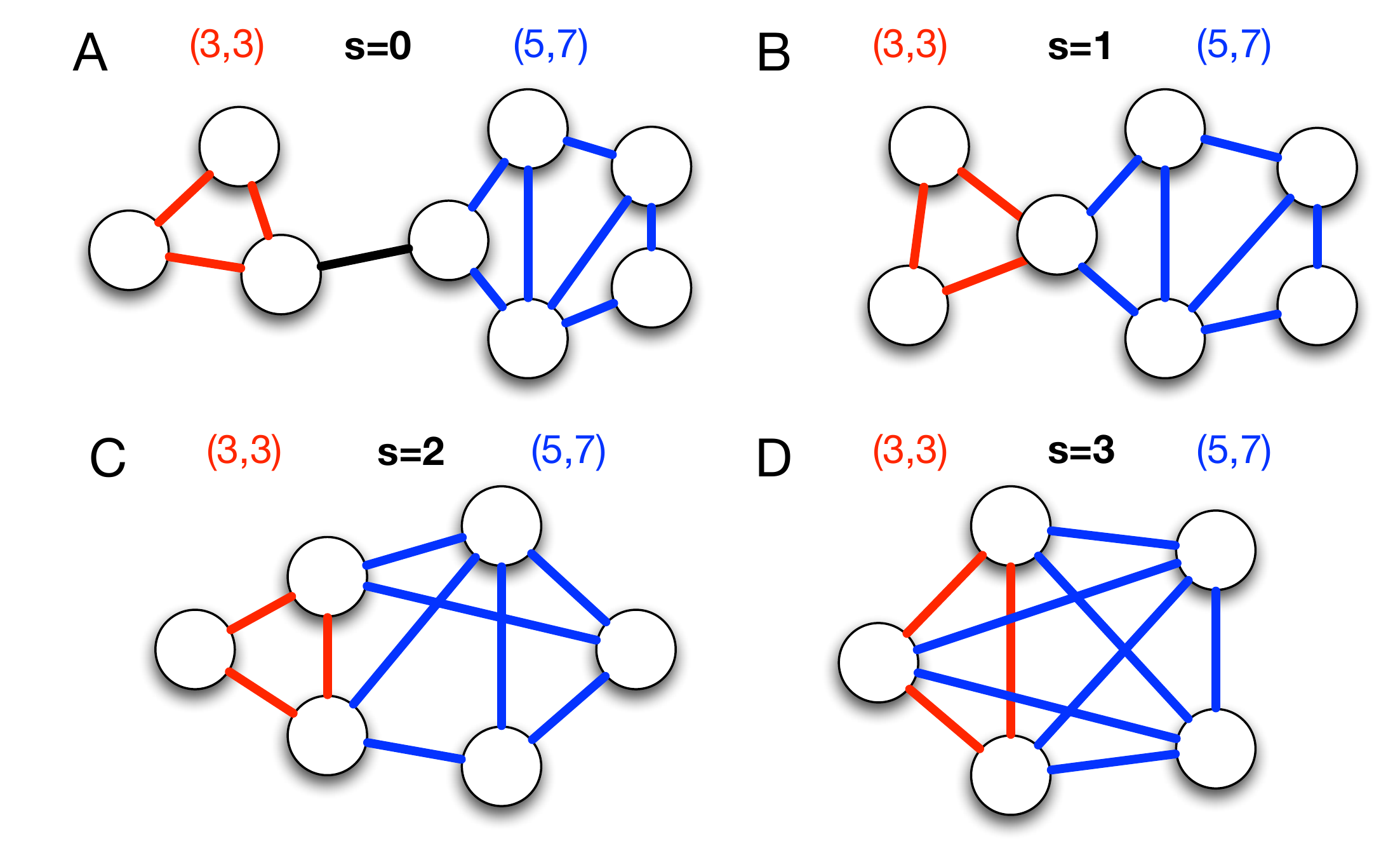}
  \caption{Schematic representations of a triangle (green) and another link
  community (red) with $n$ nodes and $m$ edges sharing $s$ nodes, (A) $s=0$, (B)
  $s=1$, (C) $s=2$, and (D) $s=3$.  Here, the number of nodes and edges of the
  other link community is set to 5 and 7, $(n,m)=(5,7)$.
  }
  \label{fig:tri_cases}
\end{figure}

As partition density has been successfully applied previously, it is natural to
ask whether it can be used as a quality function subject to direct global
optimization, as in the case of modularity. However, as we will show below,
partition density heavily suffers from its preference towards small cliques since
it measures pure density without incorporating any statistical null model.  In
this section, we examine partition density's strong preference towards small
cliques in detail.  Without loss of generality, we can imagine that there is
one triangle $T$ in a local link community $C$.  Let us assume that $T$ shares
$s$ nodes with the rest of the link community $R$ containing $n$ nodes and $m$
edges.  There are four possible choices for the value of $s$ which is shown in
Figure~\ref{fig:tri_cases}.

By definition, a partition density $D$ of the community $C$ is:
\begin{equation}
  \label{D_all}
  D = \frac{2}{M} \frac{(m+3) \left(m - n + s + 1 \right)}{(n - s + 1)(n - s + 2)}
\end{equation}
where $M$ is the total number of links in the whole network, $m+3$ and $n+3-s$ are the number of nodes and links in the community $C$, respectively.

The partition density $D_T$ and $D_R$ of the triangle $T$ and the subnetwork $R$ are
\begin{equation}  \label{D_tri}
  D_{T} = \frac{3}{M},
\end{equation}
and
\begin{equation}  \label{D_other}
  D_{R} = \frac{2}{M}\frac{ m(m - n + 1) }{ (n-1)(n-2) },
\end{equation}
respectively.

The condition where the separation of triangle $T$ is preferred can be determined by solving the following inequality:
\begin{align}
  \Delta D & = D_1 + D_2 - D \\[5pt]
  & = \displaystyle\frac{1}{M} \left( 3 + \displaystyle\frac{2m(m-n+1)}{(n-1)(n-2)} - \displaystyle\frac{2(m+3)(m+s-n+1)}{(n-s+2)(n-s+1)} \right) > 0,
  \label{eq:inequal}
\end{align}
If $\Delta D$ is negative, the triangle $T$ and its neighboring link community $R$ will merge into one link community, otherwise they prefer to be separated.

When $s=0$,
\begin{align}  \label{eq:delta_d01}
\Delta D & = \frac{1}{M} \left( 3 + (m-n+1) \left( \frac{2m}{(n-1)(n-2)} - \frac{2(m+3)}{(n+1)(n+2)} \right) \right) 
\end{align}
If $m$ is replaced with the minimum number of links between $n$ nodes,
$n-1$, $\Delta D = 3/M$, which is positive.  Because $\Delta D$ is an
increasing function of $m$, $\Delta D$ is always positive; the formation
of triangle is always preferred.

Similarly, if $s=1$,
\begin{align} \label{eq:delta_d11}
 \Delta D = \frac{1}{M}\frac{(4n-2)m^2 - (8n^2-18n+10)m + 3n^3 - 15n^2 + 24n - 12}{(n-2)(n-1)n(n+1)}.
\end{align}
Because $n>2$, the denominator $(n-2)(n-1)n(n+1)$ is positive, and the
coefficient of $m^2$, $4n-2$, is also positive.  Fixing $n$, $\Delta D$ 
is a monotonically increasing function of $m$ if $m$ is larger than
$(4n^2-9n+5)/(4n-2)$, which is smaller than the minimum of $m$, $n-1$.  
Replacing $m$ by $n-1$, $\Delta D$ is positive.  
Hence $\Delta D$ is always positive for $s=1$.

If $s=2$,
\begin{align}\label{eq:delta_d22}
 \Delta D = \frac{1}{M}\frac{4m^2 + (24-14n)m + 3n^3 - 3n^2 - 24n + 36}{(n - 2)(n - 1)n}.
\end{align}
Similar analysis shows that fixing $n$, $\Delta D$ is a monotonically increasing function of $m$ if $m$ is larger than $(7x-12)/4$, which is larger than the minimum of $m$, $n-1$. 
Replacing $m$ by $(7n-12)/4$, $\Delta D$ is positive except for the case $n=3$. 
For $n=3$, $\Delta D$ is negative when $m$ has its minimum value $n-1=2$.
$\Delta D$ keeps decreasing as $m$ increases until $m = (7n-12)/4=9/4$. 
After $m=9/4$, $\Delta D$ increases and becomes positive again when $m = 3$.
Hence $\Delta D$ is always positive except in the case $n=3$ and $m=2$.

This result clearly shows why triangles are preferred by the current definition of
partition density. It indicates that, for a given link community consisting of
more than 4 nodes, if there exists \emph{an independent triangle} that contains a node
that is not connected with the rest of nodes in the same community, the
triangle is \emph{always separated.}

\begin{figure}[ht!]
  \includegraphics[width=0.85\columnwidth]{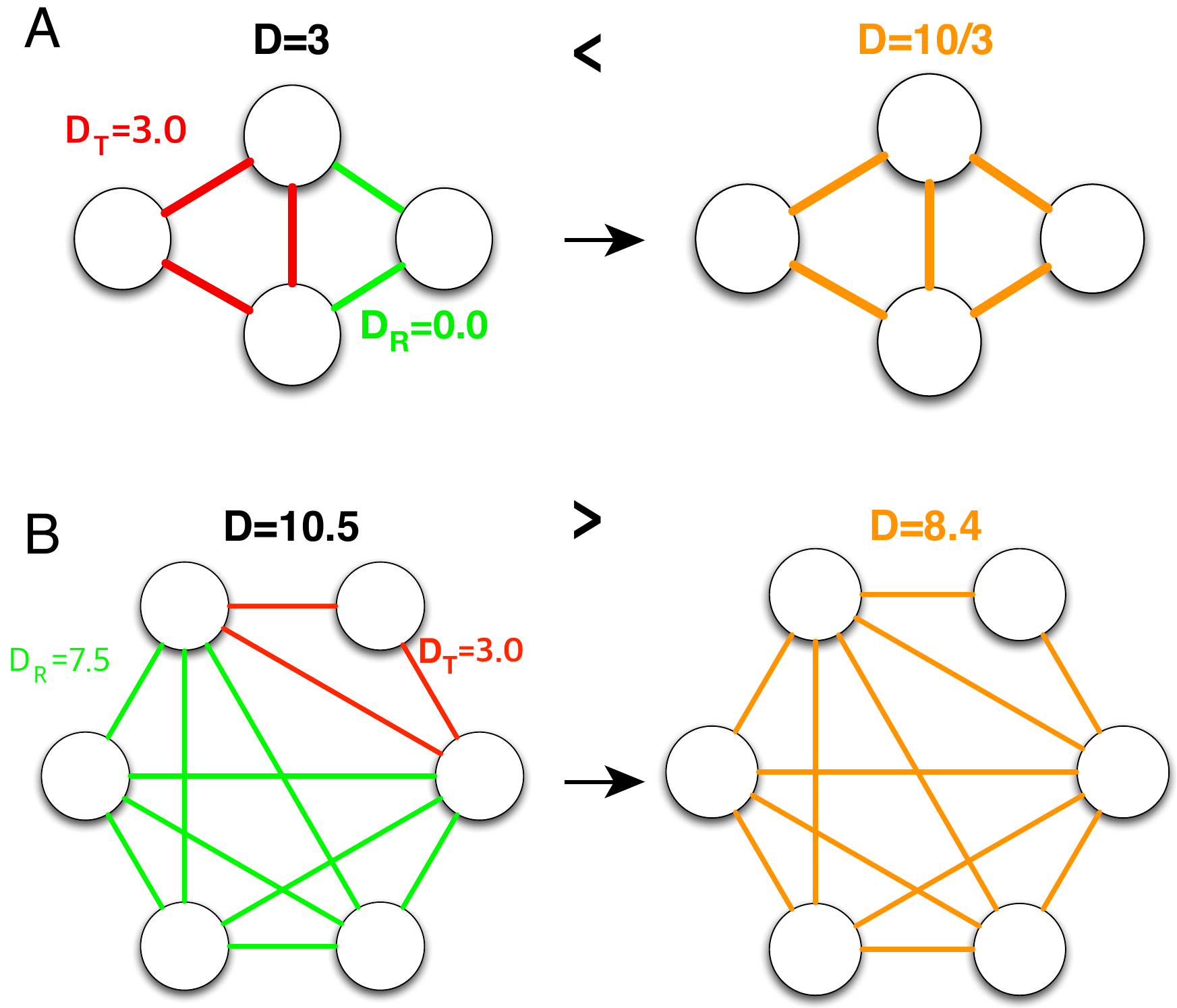}
  \caption{Examples of link communities that is (A) not separable and (B) separable when two nodes are shared between a triangle and the rest of link community, $s=2$.}
  \label{fig:z_eq_2_example}
\end{figure}

Figure~\ref{fig:z_eq_2_example} shows the examples of $s=2$ cases. 
In Figure~\ref{fig:z_eq_2_example}A, the partition density of the
green triangle is $3$, and the rest of links form a linear community
with the partition density of $0$, which results in the total
partition density of 3.  Here, the denominator $M$ in
eq.~\ref{eq:total_partition_density} is omitted since it is a
constant.  However, when the two link communities are merged, the
partition density becomes 10/3, which makes the separation of triangle
unfavorable.  In Figure~\ref{fig:z_eq_2_example}B, a link community
has more than five edges and an independent triangle.  In the right
side of Figure~\ref{fig:z_eq_2_example}B, a link community consists of
6 nodes and 12 edges and contains an independent triangle.  The 
partition density of the community is 8.4.  However, if the
independent triangle is separated, the sum of partition densities becomes
10.5, which makes the separation of triangle favorable.

\begin{figure}[ht!]
  \includegraphics[width=0.85\columnwidth]{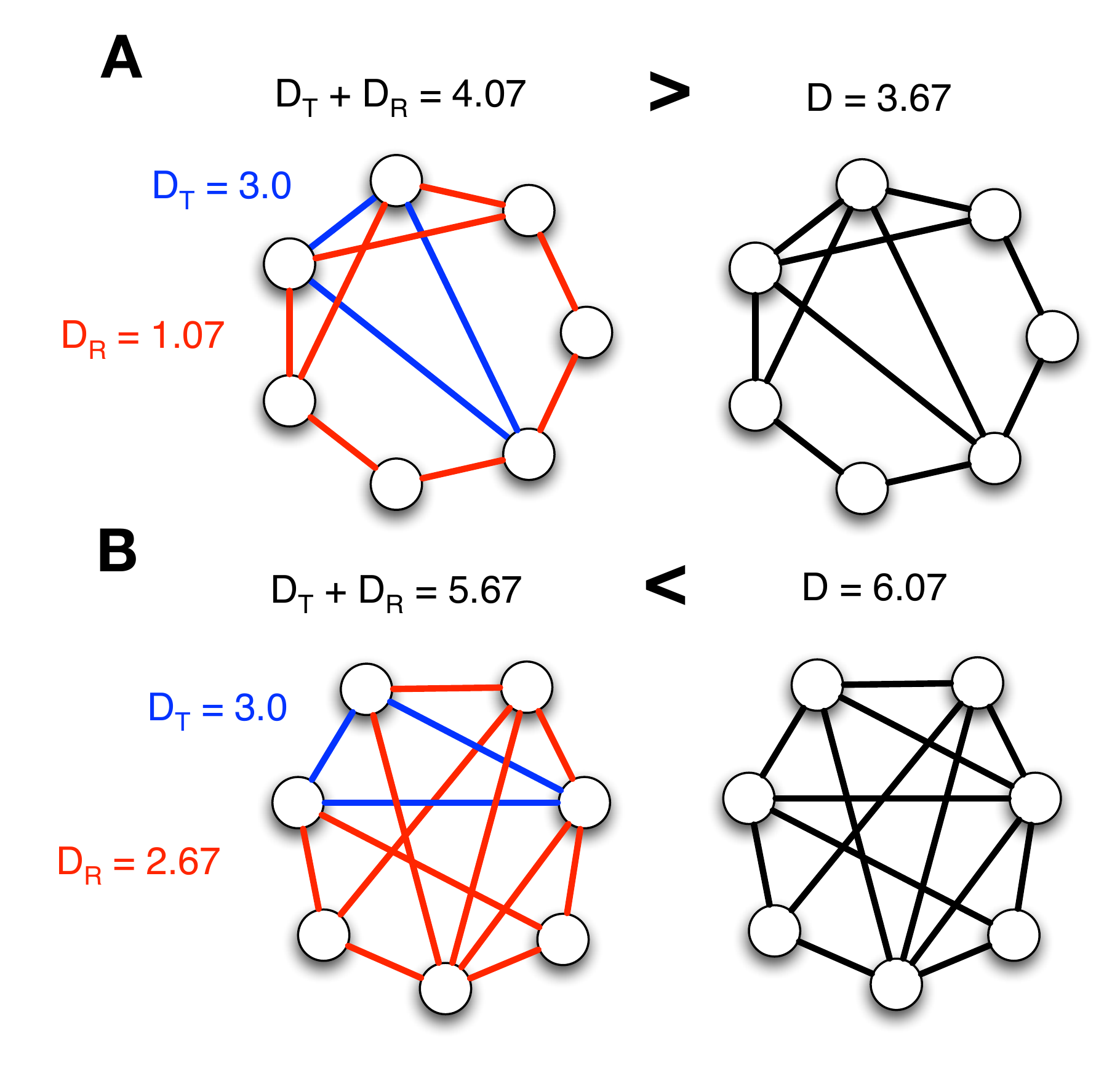}
  \caption{
  Examples of link communities when there is no separable triangle ($s=3$) where the separation of a triangle is (A) preferred and (B) not preferred.
  }
  \label{fig:z_eq_3_examples}
\end{figure}

If $s=3$, which means that there is no independent triangle in the community,
$\Delta D$ can be written as below:
\begin{align}\label{eq:delta_d22}
\Delta D = \frac{1}{M}\frac{3(n^2 - n - 4m - 6)}{(n - 1)(n - 2)}.
\end{align}
$\Delta D$ is negative if the following condition is satisfied:
\begin{equation} \label{eq:neg_D_when_z_eq_3}
  m > \frac{1}{4}( n^2 - n - 6).
\end{equation}
Thus, a merged link community $\alpha$ with $n_{\alpha}$ nodes and $m_{\alpha}$ links
is non-separable if the following condition is satisfied
\begin{equation} \label{eq:neg_D_with_z_eq_3_2}
  m_{\alpha} > \frac{1}{4} ( n_{\alpha}^{2} - n_{\alpha} - 6 ) + 3.
\end{equation}
In other words, if eq.~\ref{eq:neg_D_with_z_eq_3_2} is not satisfied,
a link community is separable although there is no independent
triangle in it.  Two examples of the case of $s=3$ are shown in
Figure~\ref{fig:z_eq_3_examples}.  The first example does not satisfy
eq.~\ref{eq:neg_D_with_z_eq_3_2} (Figure~\ref{fig:z_eq_3_examples}A).  
Thus it prefers to be separated although there is no independent 
triangle.  The partition density of
the merged link community is 3.67, while the sum of partition
densities of two separated link communities is 4.07.  On the other
hand, the second example corresponds to the case where
eq.~\ref{eq:neg_D_with_z_eq_3_2} is satisfied
(Figure~\ref{fig:z_eq_3_examples}B).  The sum of partition densities
of separated link communities, 5.67, is smaller than that of the
merged link community, 6.07.  Thus, there is no
independent triangle and the separation of triangle is not preferred.

\begin{figure}[ht!]
  \includegraphics[width=0.85\columnwidth]{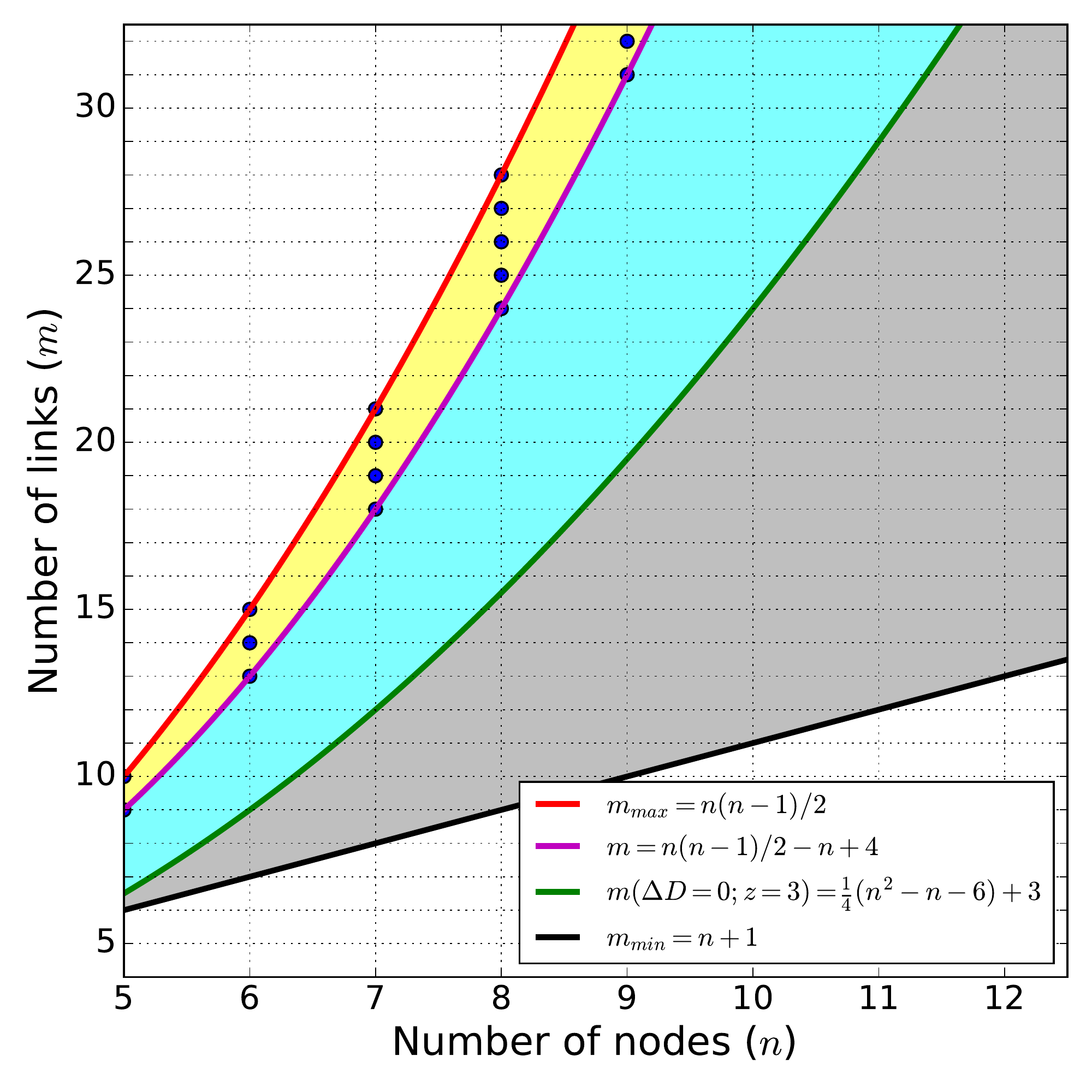}
  \caption{
    Region plot that shows the condition where the separation of triangle is not preferred (yellow), conditionally (cyan), and always preferred (gray).
    Red line represents the maximum number of link that can be formed with $n$ nodes.
    Magenta line corresponds to the minimum number of links that the separation of triangle is impossible.
    Green line represents the solution $\Delta D = 0$ when $s=3$.
    Black line represents the minimum number of links to form a link community including a triangle.
    Blue dots correspond to the conditions that a link community is non-separable.
  }
  \label{fig:triangle_region_plot}
\end{figure}

Based on these results, we can derive the condition that a link community
becomes absolutely non-separable: no independent triangle exists and
eq.~\ref{eq:neg_D_with_z_eq_3_2} is satisfied.  The maximum number of links
that has an independent triangle can be found by assuming a link community that
only one node has two direct neighbors while the rest of nodes are fully
connected to each other.  If one additional link is added in this link
community, all nodes must have at least three links, which excludes the
existence of an independent triangle.  This condition is equivalent to removing
$n-3$ links from $n$-clique, 
\begin{equation}
 \label{eq:non_separable}
  m = n(n-1)/2 - (n-3),
\end{equation}
which is always larger than eq.~\ref{eq:neg_D_with_z_eq_3_2} (Figure~\ref{fig:triangle_region_plot}).
Therefore, if a link community with $n$
nodes has more than $n(n-1)/2 - (n - 3)$ links, the community cannot contain any
independent triangle, which makes it absolutely not separable.

In summary, if a link community including 5 or more nodes is not separated 
only when it satisfies eq.~\ref{eq:neg_D_with_z_eq_3_2} and does not have an independent triangle. 
If there is an independent triangle in a link community, it is always separated.
If a link community is highly cliquish so that it satisfies 
eq.~\ref{eq:non_separable}, it is always not separated.
In conclusion, the condition
where a community will not be broken into smaller chunks under partition
density optimization is extremely limited, and the direct global optimization of
partition density will result in mostly triangles with very few larger
communities, failing to identify ``communities'' that are commonly
conceptualized. 

\section{Alternative optimization approaches}

The clear limitation of partition density as a quality function for
community detection is established. Thus, we seek alternative optimization
approaches to identify link communities. Since link community (line graph)
approach consists of (i) weighted line graph transformation~\cite{Evans2009,
Evans2010, ahn2010link, west2001introduction} and (ii) clustering in the line
graph space, we explore both the methods of weighted line graph transformation
and other quality functions that can be optimized to identify disjoint
communities in the transformed line graph. 

The main rationale behind the \emph{weighted} line graph transformation is to
mitigate the problem that each hub becomes a huge clique by penalizing the
connections between the links that are attached to the same hub.  One approach
to estimate the similarity between two links is based on preserving the
dynamics of a random walker on the links of a
network~\cite{Evans2009,Evans2010}.  In this approach, if two links are
connected via a node whose degree is $k$, the weight of the corresponding link
in the line graph becomes $1/(k-1)$.  We will call this approach a simple
normalization scheme.  Another approach is incorporating the information about
the neighbors and ignoring the contribution from the shared node, by employing
the Jaccard index~\cite{Jaccard1912, ahn2010link}. In this approach, if we
consider two links ($e_{ik}$ and $e_{jk}$) that are attached to a hub $k$, their
similarity is entirely based on the neighbors of $i$ and $j$, ignoring $k$'s
neighbors. The similarity measure is defined as follows:
\begin{equation}
 J(e_{ik}, e_{jk}) = \frac{|n(i) \cap n(j)|}{|n(i) \cup n(j)|},
 \label{eq:orig_Jaccard}
\end{equation}
where $n(i)$ is the set of the direct neighbors of node $i$ and itself~\cite{ahn2010link}.
Compared to the simple normalization scheme, two distinct pairs of links connected via the same node can have different weights using the Jaccard index.
In addition to the original Jaccard index, we introduce a normalized Jaccard index, which is the combination of the two above:
\begin{equation}
 J_{norm}(e_{ik}, e_{jk}) = \frac{1}{d(k)} \frac{|n(i) \cap n(j)|}{|n(i) \cup n(j)|},
 \label{eq:orig_Jaccard}
\end{equation}
where $d(k)$ is the degree of node $k$. This similarity measure reduces the strength of cliques formed by hubs even further by not only incorporating the local information but also penalizing large hubs. 

Once we generate line graphs based on the three aforementioned schemes, we can
perform disjoint, node community detection on those weighted line graphs. As
alternatives to partition density, we simply choose two of the most widely used
quality functions: modularity and map equation (Infomap)~\cite{Infomap}. 

\section{Experimental results}

\subsection{Conformational space annealing (CSA) algorithm}

We optimized partition density directly to show the preference of partition
density toward triangles by using the conformational space annealing algorithm
(CSA), which has been successfully applied to various global optimization
problems~\cite{lee2003unbiased, Joo2008, Joo2009,
Shin2011,lee2011novo,lee2012modularity,Sim2012,Joo2014, Joo2015}.  
We converted the CSA
implementation for modularity optimization~\cite{lee2012modularity} to optimize
partition density.  Each solution is a $N$-dimensional vector
representing the community indices of $N$ links in a network.  For local
optimization of $D$, we used a quench procedure, which accepts a move only 
when $D$ is improved, equivalent to simulated annealing at zero temperature.  
Detailed description on a general CSA algorithm can be found
elsewhere~\cite{lee1997new,lee2012modularity}.

\subsection{Partition density optimization of synthetic networks}

In this study, we used two classes of synthetic networks to show the limitation
of partition density: Girvan-Newman (GN)~\cite{girvan2002community} and
Lancichinetti-Fortunato-Radicchi (LFR)~\cite{lancichinetti2009benchmarks}
benchmark networks.  The GN benchmark network consists of 128 nodes divided
into 4 node communities of 32 nodes.  Each node is connected to the other nodes
in the same community with $Z_{in}$ links and to nodes in other modules with
$Z_{out}$ links.  Every node has 16 links in total, $Z_{in}+Z_{out}=16$.  When
$Z_{in} > 8$, each node has more connections within the community than the rest
of network and corresponds well to the four pre-defined communities.  
In the LFR network, the node degrees and
community sizes are stochastically assigned to follow a power-law distribution.
Links are stochastically connected based on a mixing parameter $\mu_{mix}$, ranging from 0 to 1.  
Each node shares a fraction of $1-\mu_{mix}$ of links with the other nodes in the 
same community, and a fraction of $\mu_{mix}$ of links with the rest of network.
Thus, a community structure becomes weaker as $\mu_{mix}$ increases, and a community
structure in a strong sense exists until $\mu_{mix} < 0.5$.  In this study, GN
networks are generated with $Z_{in}$ values ranging from 4 to 12.  LFR networks
are generated with a degree distribution that follows a power-law distribution
with an exponent of 2 ranging from 10 to 50. Community sizes are tuned to
follow a power-law distribution with an exponent of 1 and ranges from 10 to 30.

\begin{figure}[ht!]
  \includegraphics[width=0.85\textwidth]{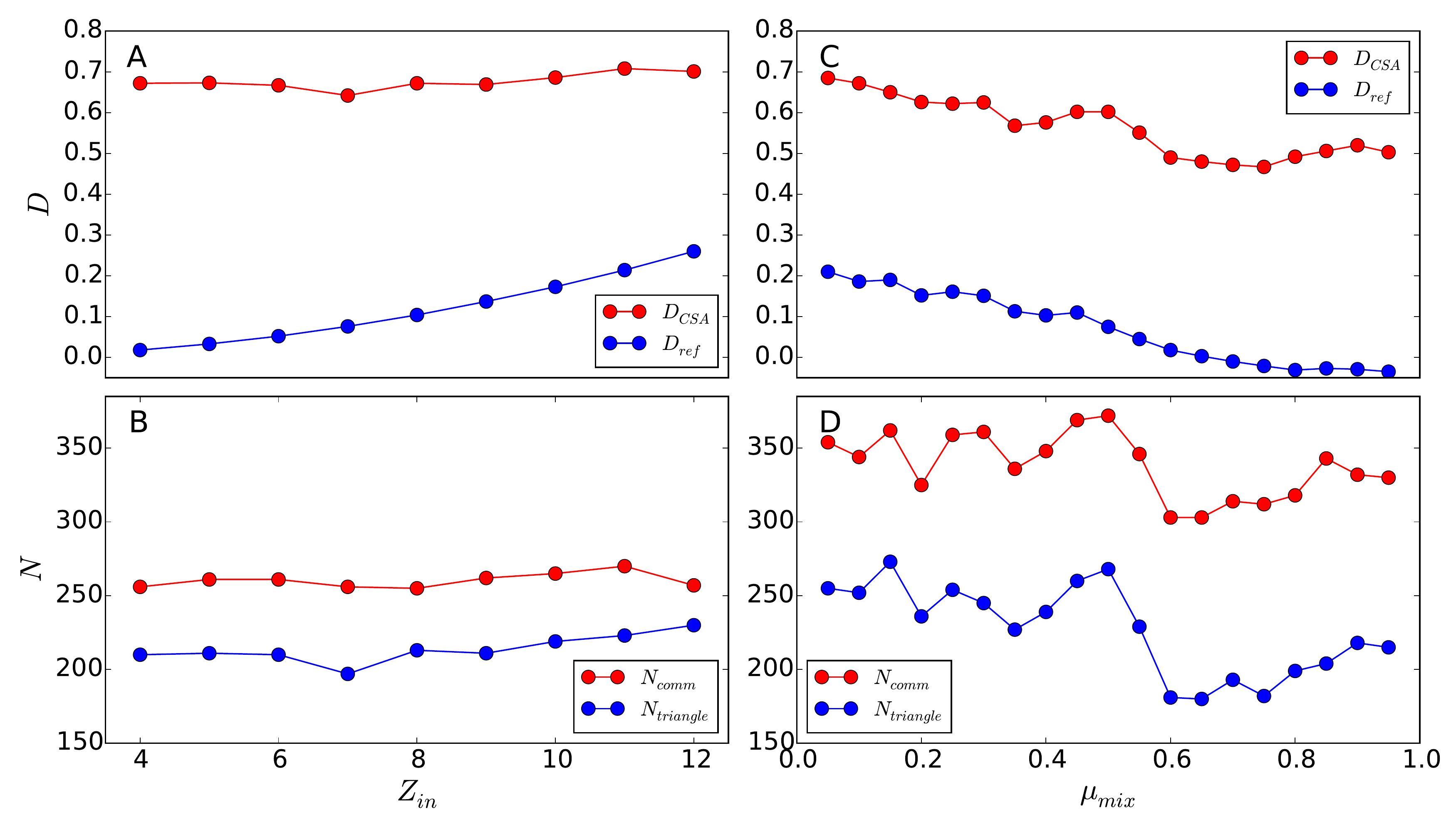}
  \caption{Optimized partition density and the estimated number of communities by optimization of partition 
  density on the GN and LFR benchmark networks.
  Subplot A and C plot the optimized ($D_{CSA}$) and reference ($D_{ref}$) partition densities versus $Z_{in}$ and $\mu_{mix}$ values.
  Subplot B and D plot the numbers of all identified link communities ($N_{comm}$) 
  and triangles ($N_{triangle}$) versus $Z_{in}$ and $\mu_{mix}$ values.}
  \label{fig:gn_net_summary}
\end{figure}

For the GN networks with different $Z_{in}$ values ranging from 4 to 12, we
compared optimized $D$ ($D_{opt}$) values by CSA with the reference $D$ ($D_{ref}$) value, which is
obtained from the pre-defined node-community structure.  To calculate the
$D_{ref}$, all intra-node-community edges of a node-community are
considered as the same link community and inter-node-community edges are
ignored.  For all GN networks, the $D_{opt}$ values are
much higher than the $D_{ref}$ values (Figure~\ref{fig:gn_net_summary}A).
The $D_{opt}$ values are almost identical for all GN networks, around 0.7,
while the $D_{ref}$ value monotonically increases from 0.03 to 0.23 as the
community structure of GN network strengthens.  We also counted the numbers of
triangles and all link communities from the CSA results
(Figure~\ref{fig:gn_net_summary}B).  For all the GN networks, around 260 link
communities are detected via $D$-optimization and, among them, around 220 link
communities are triangles on average.  In addition, it is noticeable that the
number of triangle increases as $Z_{in}$ increases, which suggests that highly
modular networks may suffer more from this drawback of $D$.  These results show
that the global optimization of $D$ leads to a significantly different community
structure from the reference community due to the triangle preference of $D$.

We performed a similar benchmark using LFR networks with different mixing
probabilities.  Overall, the benchmark on the LFR networks shows a
qualitatively similar results with those on the GN networks.  A comparison of
$D_{opt}$ and $D_{ref}$ values demonstrates that there is a large gap
between two values regardless of $\mu_{mix}$, and both $D$ values decrease as
networks become less modular, a larger $\mu_{mix}$
(Figure~\ref{fig:gn_net_summary}C).  The inverse correlation between $D$ and
$\mu_{mix}$ shows that $D$ is correlated with the degree of modularity.
However, as shown in the GN networks, community structures with high $D$ values
does not correspond to the true community structure.  From
Figure~\ref{fig:gn_net_summary}D, it can be identified that about 2/3 of
detected link communities via $D$-optimization are triangles, and more
triangles are detected in the networks with a strong sense of community,
$\mu_{mix} < 0.5$, than the networks without community, $\mu_{mix} > 0.5$.

\subsection{Partition density optimization of real-world networks}

\begin{figure}[ht!]
  \includegraphics[width=0.85\textwidth]{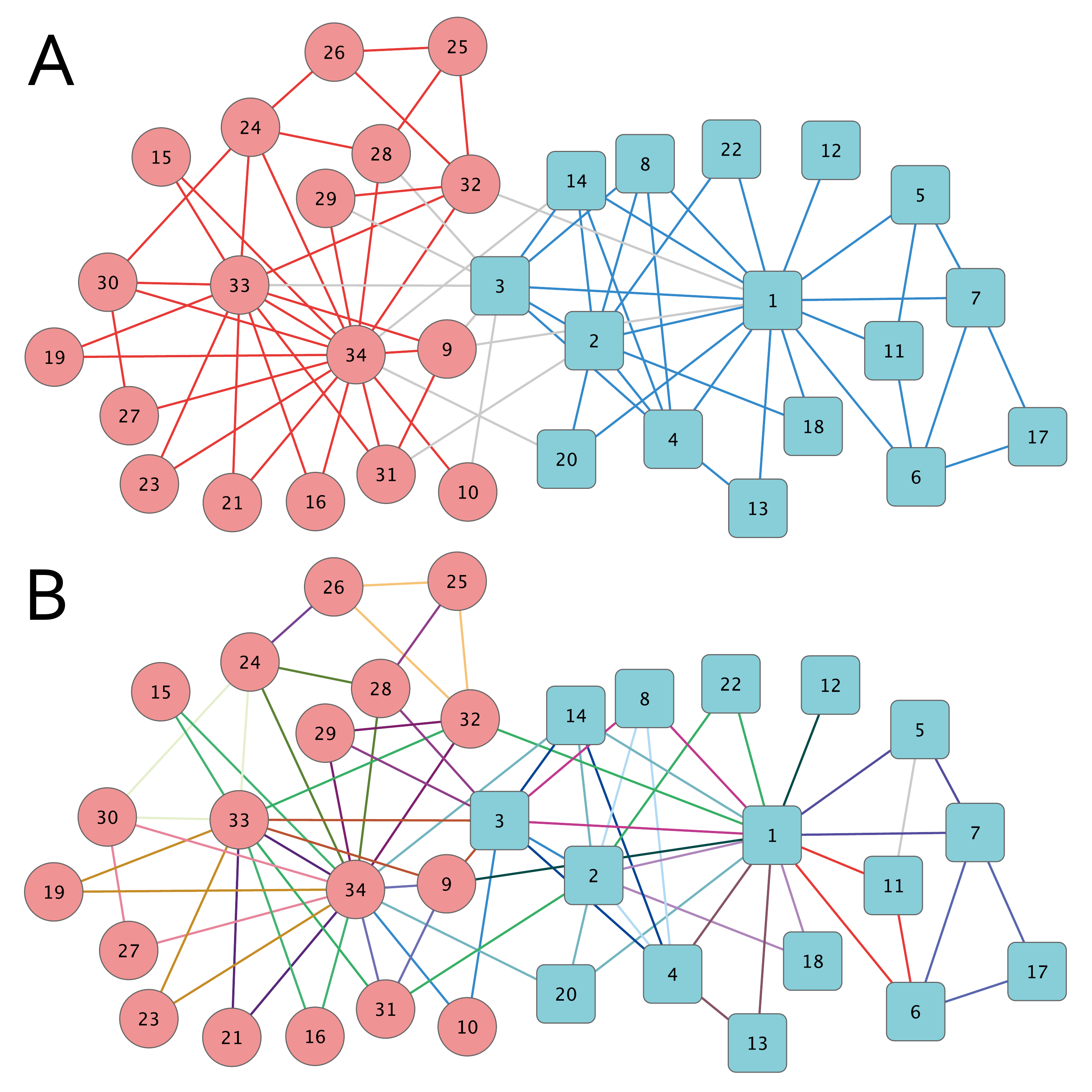}
  \caption{A comparison of real grouping and communities found by optimization of partition density on karate network. 
  The node shape and colors represent the real groupings and the edges are colored by link communities.}
  \label{fig:karate}
\end{figure}

We also performed $D$-optimization of several popular real-world
benchmark networks (Table.~\ref{tab:CSA_result_of_real_networks}).
For all real-world benchmark networks, more than half of detected link
communities by $D$-optimization are triangles, which indicates that
the inverse resolution limit of partition density is universal.  A
comparison of the real groupings and the link communities obtained by
$D$-optimization of karate network is shown in
Figure~\ref{fig:karate}.  In the real grouping, the karate network is
divided into two and the $D$ value of the corresponding link community
is 0.128 (Figure~\ref{fig:karate}A).  However, the CSA result with
$D=0.683$ divides the karate network into 25 link communities, which
includes 18 triangles and 4 unclassified edges
(Figure~\ref{fig:karate}B).  There are two largest link communities
consist of 6 edges connecting nodes 1-2-14-20-34 and 1-22-2-31-33-32,
which mainly consist of hub nodes.

The results of the synthetic benchmark networks and the karate network give us a hint of how
$D$-optimization divides a network.  $D$-optimization finds as many
cliques as possible because their local partition density, $D_\alpha$, is 1.
Since the smallest clique is a triangle, most of edges are grouped as a
triangle.  After removing the detected cliques from the network,
$D$-optimization finds densely connected groups of the rest of edges, which are
likely to be loosely connected groups of hub nodes due to their large degrees.
The rest of disconnected edges remain unclassified, \emph{e.g.} edges
connecting 1-12 and 5-11 in Figure~\ref{fig:karate}B.

\begin{table}[ht!]
  \caption{The number of triangles and the total number of link communities 
  of real world networks obtained with the optimization of partition density.}
  \centering\scriptsize
  \begin{tabular}{c || c | c} \hline\hline
    Dataset           & \# Triangle communities & \# Communities \\ \hline
    Karate            & 18   & 25   \\
    Dolphin           & 30   & 51   \\
    Lesmis            & 26   & 44   \\
    Political books   & 83   & 120  \\
    Football          & 109  & 168  \\
    Netscience\_main  & 98   & 200  \\
    \emph{C. elegans} & 297  & 512  \\
    Jazz              & 562  & 772  \\
    \emph{E. coli}    & 1466 & 2184 \\ \hline\hline
    \label{tab:CSA_result_of_real_networks}
  \end{tabular}
\end{table}


\subsection{Modularity optimization of weighted line graphs}

We used LFR benchmark networks for overlapping communities with 1000 nodes,
mixing probabilities ($\mu_{mix}$) of 0.1 and 0.3, and the numbers of memberships 
of an overlapping node (om) of 2 and 4~\cite{lancichinetti2009benchmarks}.  With each
parameter set, ten different networks were generated and tested.  The qualities
of community detection results were evaluated by calculating the 
normalized mutual information (NMI) values for
overlapping communities~\cite{Lancichinetti2009} between the obtained
communities and the reference communities.  Recently, it was reported that NMI
is affected by the finite size of a network and the finite number of detected
communities~\cite{Zhang2015}.  To remove this artifact, the NMI values are
adjusted by subtracting the average NMI values of randomly shuffled
communities while preserving the number of communities.

\begin{figure}[ht!]
  \includegraphics[width=0.9\textwidth]{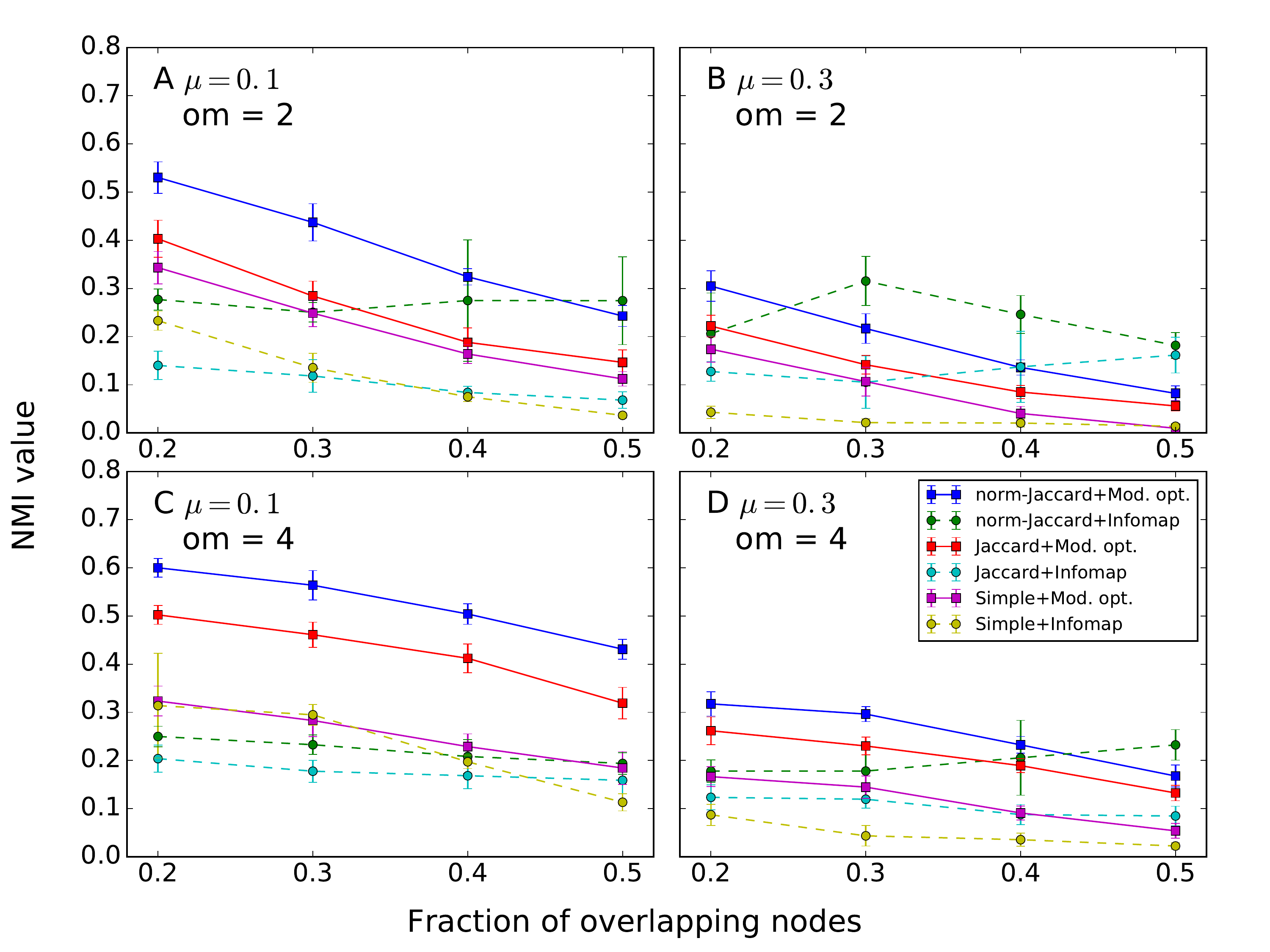}
  \caption{
  Normalized mutual information values of the community detection results using 
  the weighted line graphs of the LFR benchmark networks with overlapping communities.
  The line graphs are generated with three different weighting schemes: 
  the simple normalization scheme (Simple), the Jaccard index (Jaccard), and the normalized Jaccard index (norm-Jaccard).
  Communities of line graphs are detected by optimizing modularity (Mod. opt.) and Infomap.
  $\mu$ is a mixing parameter and om is the number of memberships of an overlapping node (om).
  }
  \label{fig:nmi_lfr_overlapping}
\end{figure}

The benchmark results clearly show that performing community detection
using the line graphs generated with the normalized Jaccard index
yields the highest NMI values in most cases
(Figure~\ref{fig:nmi_lfr_overlapping}).  With the modularity
optimization method, the line-graphs generated with the normalized
Jaccard index lead to higher NMI values for all parameter sets tested.
With the the Infomap method, higher NMI values were consistently
obtained with the normalize Jaccard index except two cases when line
graphs are generated with the simple normalization scheme and
$\mu=0.1$, om=4, and the fraction of overlapping nodes are less than
0.4.  In summary, these results indicate that detecting disjoint
communities of the line graphs generated with the normalized
Jaccard index leads to more meaningful link communities.

\section{Discussion and Conclusion}

In this study, we showed that partition density suffers from the strong
preference towards small cliques; identifying triangles as separate link communities
is preferred in most possible scenarios.  Direct global optimization of partition
density of the synthetic and the real-world networks resulted in a huge number
of triangles.
We showed that a triangle contains a node that is connected only 
with the other two nodes, it always prefers to be separated. 
The only exception is when 4 nodes are connected with 5 links.

One of the reasons for the preference to a triangle is that a difference in
local partition density $D_\alpha$ between a triangle and larger cliques or
cliquish link communities becomes marginal as a network becomes larger.  By
definition, a decrease in $D_\alpha$ of a large link community due to a
separation of a triangle becomes smaller as $n_\alpha$ increases
(eq.~\ref{eq:local_partition_density}).  However, $D_\alpha$ of a separated
triangle is 1.0, which can be larger enough to compensate the decreased
$D_\alpha$ of the initial link community.  Our result raises further questions:
how should we handle triangles? Is it more meaningful than a larger cliquish
link community?  Although a triangle is a clique, it may be too small to
extract meaningful information from it and to reduce the complexity of a
network efficiently.  Thus, a criterion to compare the significance of a
triangle and larger cliquish link communities may be necessary.

Considering the strong bias of the partition density how could it work as a
quality function for the link clustering method~\cite{ahn2010link}?  First, a
hierarchical clustering was performed in an agglomerative way to generate the
dendrogram of links and detect the community structure of a network based on a
threshold that maximizes partition density.  With this approach, formation of
triangles is suppressed because clustering is carried out in a way that the size
of a cluster increases only by merging the most similar pair of nodes first,
imposing strong constraints on the community structures.  Second, the
heterogeneity of a network might play an important role.  If the degree
distribution of nodes follows an uniform or a Gaussian distribution, many nodes
may have similar numbers of links, direct neighbors, which makes most pairs of
links have similar similarities.  If this is the case, many triangles may have
formed due to a high degeneracy of priorities of links for merging.  However,
many real-world networks are known to be scale-free networks whose degree
distributions are highly heterogeneous.  The heterogeneity of connectivity
leads to a heterogeneous distribution of link similarities, which leads to the
formation of the hierarchical organization of link
communities~\cite{ahn2010link}.

As an alternative optimization approach to the partition density, we explored
the possibilities of directly optimizing other quality functions combined with
several versions of weighted line graph transformations.  In the previous scheme
suggested by Evans and Lambiotte, the weight of a link in a line-graph is set
to the inverse of the degree of a common node~\cite{Evans2009,Evans2010}.  This
approach preserves the dynamics of random walkers.  However, information on the
similarities between links of an initial network is not reflected explicitly.
Thus, this scheme may not be useful when one wants to investigate the
relationships between links using a line-graph representation. 
In the previous link clustering study~\cite{ahn2010link}, the Jaccard index was used.
We suggested the normalized Jaccard index, which combines these two previous 
approaches.  Modularity optimization of the line-graphs
generated with the normalized Jaccard index resulted in more accurate community
structures than those with the Jaccard index or the simple normalization scheme
alone, prompting further investigation into good quality functions and
weighting schemes for line-graph based link community detection.

\begin{acknowledgments}

YYA thanks for the support from Microsoft Research Faculty Fellowship.

\end{acknowledgments}

\bibliography{pd}
\end{document}